\documentclass{xeus}
\usepackage{graphics}
\usepackage{psfig}
\begin{document}
\onecolumn
\title{High resolution spectroscopy of nearby AGN}
\author{J.S. Kaastra\inst{1}}
\institute{SRON National Institute for Space Research, Sorbonnelaan 2,
3584 CA Utrecht, The Netherlands}
\maketitle

\begin{abstract}

In this paper the potential of high resolution spectroscopy of nearby AGN with
XEUS is discussed.  The focus is upon the energy resolution that is needed in
order to disentangle the different spectral components.  It is shown that there
is an urgent need for high spectral resolution, and that a spectral resolution
of 1~eV, if possible, leads to a significant increase in diagnostic power as
compared to 2~eV resolution.

\end{abstract}

\section{Introduction}

Since the launch of Chandra and XMM-Newton, with their high resolution
spectrometers, our insight into the astrophysics of Active Galactic Nuclei (AGN)
has changed dramatically.  It is now possible to study in detail the geometry,
dynamics and physical state of the warm absorber, as well as the underlying
continuum spectrum, including the exciting possibility of relativistic emission
lines.  While with XMM-Newton and Chandra high resolution, high signal-to-noise
ratio spectra of the brightest (and in general the nearest) AGN can be taken,
the large resolving power and effective area of XEUS will allow us to study AGN
spectra out to large redshifts or low intrinsic luminosities.  It will also
allow time-resolved spectroscopy of the most rapidly varying AGN.  A detailed
understanding of the astrophysics and spectral signatures of the nearest AGN is
an absolute requirement in order to understand the properties of the most
distant AGN that will be observed by XEUS, which will have much noisier spectra.
In this contribution  spectral simulations  are presented that show the
potential and limitations of AGN spectroscopy with XEUS.

\section{Spectral simulations}

I have performed a set of spectral simulations of a bright Seyfert 1 galaxy as
will be observed with XEUS.  For the effective area of the instrument I took the
area as foreseen in the final configuration of XEUS (taken from the XEUS web
page).

For the spectral resolution a parameterization of what currently is thought to
be feasible was taken:  a constant full width at half maximum (FWHM) of 2~eV
below 1~keV, 10~eV above 14~keV and in between a linear interpolation.  A
spectral response matrix was generated with a Gaussian energy resolution
consistent with the FWHM and effective area mentioned above.  All spectral
simulations were done using the SPEX package version 2.0 (Kaastra et al.
\cite{kaastra2002spex}).

For the Seyfert 1 spectrum I took a prototype of a bright and nearby AGN:
NGC~5548.  This Seyfert~1 galaxy was the first to be observed at high spectral
resolution (Kaastra et al.  \cite{kaastra2000}).  It has a moderate redshift of
0.017, and a low Galactic column density.  The spectral model I used is
described below, and is based upon the modeling of the Chandra LETGS data
(Kaastra et al.  \cite{kaastra2002a}) for most energies, supplemented by the
Fe-K line emission as modeled by Yaqoob et al.  (\cite{yaqoob}) based upon the
Chandra HETGS data.

The ingredients of our spectral model are:

\begin{itemize}
\item a power law continuum
\item a modified blackbody spectrum, describing the accretion disk
    continuum at low energies 
 \item a warm absorber, consisting of three
ionization and five velocity components, with a  range of
 $-$160 to $-$1060~km/s outflow velocity
 \item narrow forbidden emission lines from \ion{O}{vii} and \ion{Ne}{ix}
 \item a narrow Fe-K line component at 6.4~keV
 \item a relativistic Fe-K line
 \item weak relativistic \ion{O}{viii} and \ion{N}{vii} Ly$\alpha$ lines
 \item a reflection component at high energies
\end{itemize}

\begin{figure}
\centerline{\psfig{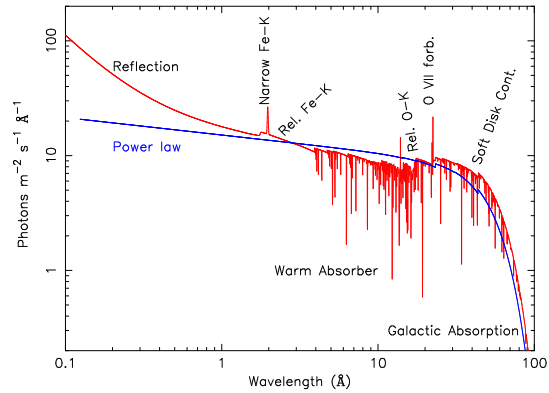}}
\caption[]{Model spectrum at high spectral resolution
for NGC~5548, with the different spectral
components as indicated. The blue line indicates the underlying power law
component. Note that nowhere in the spectrum the total model spectrum
looks similar to the underlying power law!}
\label{fig:modspec}
\end{figure}

\begin{figure}
\centerline{\psfig{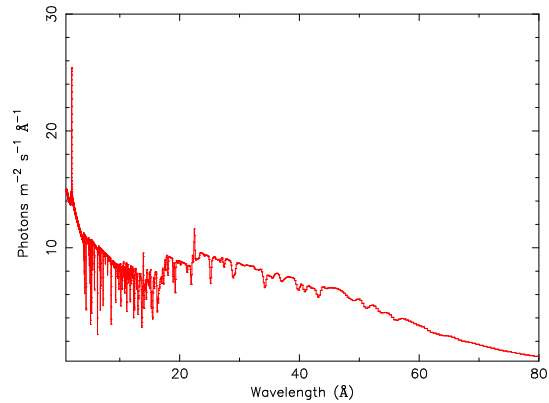}}
\caption[]{Model spectrum for NGC~5548, 
folded with the energy resolution of XEUS.}
\label{fig:figtotang}
\end{figure}

The model photon spectrum corresponding to the above model, as would be seen
with very high energy resolution, is shown in Fig.~\ref{fig:modspec}.  The model
spectrum, but now folded with the XEUS spectral resolution, is shown in
Fig.~\ref{fig:figtotang}.  Note that nowhere in the model spectrum there exists
an energy range where the spectrum agrees fully with the underlying power law
component.  At high energies there is the excess of the reflection component,
around 6~keV the broad iron line component causes excess flux, then at lower
energies the warm absorber reduces the observed flux, followed by excess
emission at the lowest energies, due to emission from the accretion disk,
including the modified blackbody as well as any possible relativistic Ly$\alpha$
lines from oxygen and nitrogen.

Thus, in order to derive the underlying AGN continuum or any of the other
spectral components, the spectrum must be fitted over the entire wavelength
range, taking into account all contributions.  It is also seen that due to the
fact that XEUS will use non-dispersive X-ray detectors, the spectral resolution
at low energies is relatively poor and not all the spectral lines can be
resolved (compare Fig.~\ref{fig:modspec} with Fig.~\ref{fig:figtotang}).  I come
back later to this point.

\section{Finding relativistic lines}

The presence of relativistic emission lines from \ion{O}{viii} and \ion{N}{vii}
Ly$\alpha$ has been discussed first by Branduardi-Raymont et al.
(\cite{branduardi}) based upon XMM-Newton RGS observations of the narrow line
Seyfert 1 galaxies MCG~--6-30-15 and Mrk~766.  In these two galaxies the
relativistic lines are very strong, although another group disputes this result
(Lee et al.  \cite{lee}).  The flux level in both sources is quite strong, so
statistics is not really the problem, but the debate depends heavily upon
disentangling the different spectral components.  A stronger warm absorber would
produce deeper oxygen absorption edges, and hence would leave less room for a
relativistic oxygen emission line.  But in order to get a reliable estimate of
the depth of the oxygen absorption edges, which are strongly contaminated by
several other absorption lines and edges, a detailed modeling of the absorption
lines is needed, and for that purpose high spectral resolution is a necessary
condition.

\begin{figure}
\centerline{\psfig{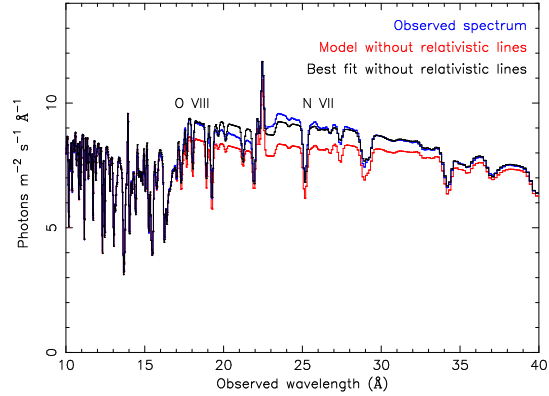}}
\caption[]{Model spectrum of NGC~5548 with the XEUS resolution near the
region of the relativistic oxygen and nitrogen lines. The blue curve
is the simulated spectrum for 40~ks integration time (error bars are very
small and invisible); the red curve is the same model spectrum but with the
relativistic lines left out; the black curve is the best fit model
for a model that does not take into account the relativistic lines.}
\label{fig:fig_nolaor}
\end{figure}

In NGC~5548 the relativistic lines are much weaker, as is shown in 
Fig.~\ref{fig:fig_nolaor}. It is seen from this figure that the maximum
amplitude of these lines is at most $\sim$~15~\%, around 24~\AA. It
is also seen that when the relativistic lines are not taken into account, 
the resulting spectral fit agrees quite well with the observed spectrum.
The differences in the continuum are less than a few percent, in a region
where very likely also instrumental neutral oxygen edges will be present.
This therefore poses severe constraints on the calibration of the effective
area, which must be better than a percent over broad energy ranges.

A way out is of course by looking to the spectral lines in order to
constrain the depth of the absorption edges. But for that
purpose the highest possible spectral resolution is needed.

\section{Complexity of the warm absorber}

\begin{figure}
\centerline{\psfig{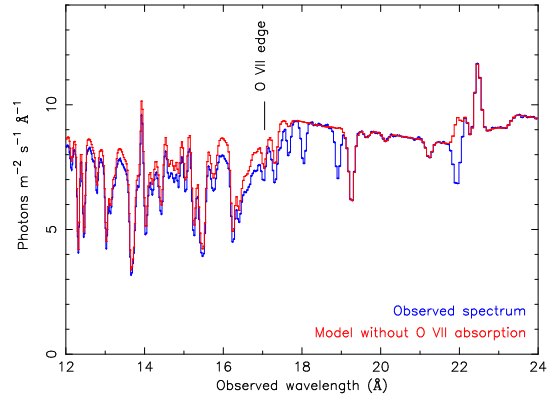}}
\caption[]{Blue line: model spectrum of NGC~5548 with the energy resolution
of XEUS. Red curve: same model, but with the \ion{O}{vii} column
density put to zero.}
\label{fig:fig_noo7}
\end{figure}

The complexity of the warm absorber is illustrated in Fig.~\ref{fig:fig_noo7}.
From the difference of the model with and without the \ion{O}{vii} it is seen
that in the observed photon spectrum the K-edge of \ion{O}{vii}
cannot be distinguished, due to strong blending by several weak inner shell
iron lines (see below). The apparent "edge" is rather smooth, and moreover
from the jump in the continuum between 16~\AA\ and 18~\AA\ it is not
possible to measure directly the depth of the oxygen edge: the effective
depth of the "edge" is significantly deeper than just the contribution
from \ion{O}{vii} alone. Again, high resolution is needed in order to
estimate accurately the depth of the \ion{O}{vii} edge, based upon the
equivalent width ratios of the absorption lines from the same ion. Due to
blending, which is in particular important at low spectral resolution, but
which cannot always be avoided even at high energy resolution due to
the finite velocity width of the absorption complexes, it is important to
observe several lines of the same ion.

\begin{figure}
\centerline{\psfig{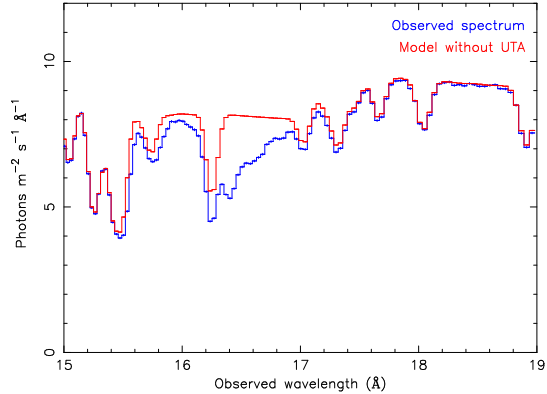}}
\caption[]{Blue curve: model spectrum of NGC~5548
with the energy resolution of XEUS; red curve: same model, but with the column
densities of the Fe-M ions put to zero.}
\label{fig:no_uta}
\end{figure}

One of the reasons for the complexity of the spectrum near the \ion{O}{vii}
edge is demonstrated in Fig.~\ref{fig:no_uta}. In this range of the spectrum
there are many weak inner-shell absorption lines due to lowly ionized iron
(in our model \ion{Fe}{ix}--\ion{Fe}{xvi}). These lines were first
recognized by Sako et al. (\cite{sako}) in their analysis of the quasar
IRAS~13349+2438. It is seen that these lines constitute a broad, unresolved
blend. They are very important from a diagnostic point of view,
since they measure directly the strength of the less ionized material
in the warm absorber, for ionization parameters that yield otherwise
only spectral lines mostly in the inaccesible (E)UV range.

\section{Why high spectral resolution is needed}

\begin{figure}
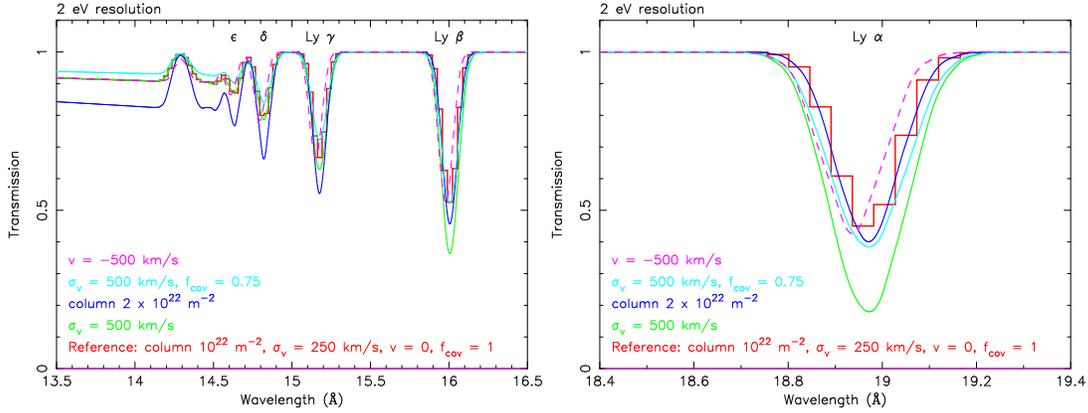

\centerline{\hbox{
\psfig{figure=jelle_kaastra_fig6a.ps,angle=-90,width=2.8in}
\psfig{figure=jelle_kaastra_fig6b.ps,angle=-90,width=2.8in}}}
\caption[]{Different models for the transmission of a slab composed of
pure \ion{O}{viii}. The model spectra are folded through an instrument with
an energy resolution of 2~eV (as is used for the present XEUS simulations).
Right panel: the Ly$\alpha$ line; Left panel: the higher Lyman lines and the
\ion{O}{viii} edge. The reference model (red curve) has an \ion{O}{viii} column
of $10^{22}$~m$^{-2}$, a velocity dispersion of 250~km/s, an outflow velocity
of 0~km/s and a covering factor of unity. The other curves have the same
parameters except the ones indicated in the caption.}
\label{fig:figo8}
\end{figure}

High spectral resolution is important in order to determine the spectral line
parameters of the warm absorber and thereby the continuum spectrum.  This is
illustrated in Fig.~\ref{fig:figo8}.  This figure simulates the transmission of
a slab composed of pure \ion{O}{viii}.  For other ions the effects illustrated
in this figure are qualitatively similar.  The model spectrum has been convolved
with the currently adopted XEUS spectral resolution of 2~eV.

For the reference model (\ion{O}{viii} column $10^{22}$~m$^{-2}$, velocity
dispersion $\sigma_{\mathrm v}$= 250~km/s, outflow velocity $v$ = 0~km/s and
covering factor $f_{\mathrm{cov}}$ = 1), the Ly$\alpha$, Ly$\beta$ and
Ly$\gamma$ lines are strongly saturated, and hence doubling the column density
does not change these line profiles significantly.  Note that for this velocity
dispersion, the intrinsic width of the Ly$\alpha$ line is about 1.2~eV and hence
unresolved with the 2~eV instrumental resolution.  In many Seyfert galaxies the
intrinsic line width can be even smaller than the value used here.  The
difference in column density can only be seen for the higher order lines
(Ly$\delta$ and higher) but these lines start becoming blended with each other
(in the model, I took all lines up to $n=10$ into account).  And of course the
difference in column density can be seen near the continuum edge, but as I have
shown before the continuum edge is often very hard to measure due to severe
contamination by various spectral lines and line blends, and due to the unknown
underlying continuum spectrum.

Measuring the intrinsic velocity broadening $\sigma_{\mathrm v}$ appears at a
first glance more promising.  By doubling $\sigma_{\mathrm v}$ to 500~km/s, the
Ly$\alpha$ and Ly$\beta$ lines become significantly deeper, while the higher
lines of the series become less deep, because they are smeared out.  The
Ly$\alpha$ and Ly$\beta$ lines become deeper because their (equivalent) width
doubles and now becomes comparable to the instrumental resolution.  However it
should be noted that the intensity at the deepest point of the {\it observed}
Ly$\alpha$ line is still not zero, despite the fact that the line core in the
original {\it model} spectrum is completely black.  This is due to the limited
spectral resolution of the instrument.

This, in fact, causes that the apparent sensitivity for changes in
$\sigma_{\mathrm v}$ is partially an illusion, as is illustrated with the light
blue curve, where the same enhanced $\sigma_{\mathrm v}$ of 500~km/s was adopted
but now with a covering factor of only 75~\%.  Now the observed line profiles
are very similar to the original reference model with $\sigma_{\mathrm v}$ of
250~km/s and full covering factor!  The difference would be most clearly visible
in the continuum edge, where the transmission differs by 25~\%, but as was
pointed out before these continuum edges are difficult to measure accurately.

Finally, Fig.~\ref{fig:figo8} also shows the sensitivity to measure outflow
velocities.  Outflow velocities of 500~km/s can be measured by using line
centroiding.  For the Ly$\alpha$ line, this velocity corresponds to a 1~eV
shift, so can be easily detected.  Note however that in many AGN the absorption
lines are composed of several velocity components, which are separated by
velocity differences on the hundred km/s velocity scale.  Each velocity
component may have its own ionization structure, hence what is observed as a
single broad absorption line is actually composed of a complex set of line
components, hence any gain in spectral resolution is extremely useful.

Needless to say that for cosmologically redshifted AGN the situation becomes
even more problematic, due to the fact that the lines shift to lower energies
with poorer instrumental resolution.

\begin{figure}
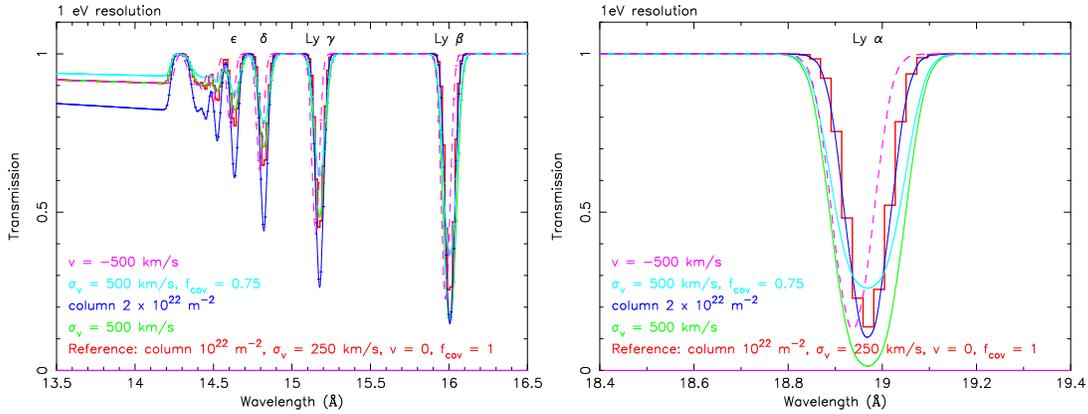

\centerline{\hbox{
\psfig{figure=jelle_kaastra_fig7a.ps,angle=-90,width=2.8in}
\psfig{figure=jelle_kaastra_fig7b.ps,angle=-90,width=2.8in}}}
\caption[]{Same as Fig.~\ref{fig:figo8}, but instead a 1~eV instrumental
resolution.}
\label{fig:figo81}
\end{figure}

Therefore I have also made a set of calculations for the line profiles with a
spectral resolution of 1~eV instead of 2~eV (Fig.~\ref{fig:figo81}).  Such a
factor of two improvement of the spectral resolution does not seem impossible
from a technical point of view, in particular if the spectral resolution is
optimized for these lower energies (below 2~keV).  Improving the spectral
resolution by a factor of two also enhances the sensitivity for weak lines with
a factor of two, without even increasing the effective area!  Comparing both
figures, it is seen that the observed lines now appear much deeper, and hence
the accurate measurement of covering factors can be done more easily.  The
differences between the various cases shown in the figure are much more
pronounced.

\begin{figure}
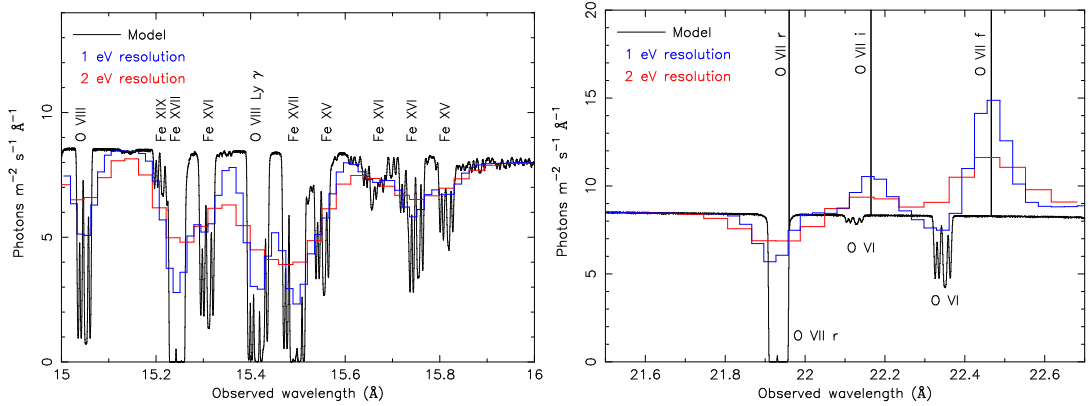

\centerline{\hbox{
\psfig{figure=jelle_kaastra_fig8a.ps,angle=-90,width=2.8in}
\psfig{figure=jelle_kaastra_fig8b.ps,angle=-90,width=2.8in}}}
\caption[]{Model spectrum at high resolution (black line), convolved with
an instrumental resolution of 1~eV (blue line) and 2~eV (red line). Left
panel: region near 15~\AA\ containing \ion{O}{viii}, \ion{Fe}{xvii}
and other important diagnostic lines; right panel: region near the
\ion{O}{vii} triplet.}
\label{fig:figreso}
\end{figure}

Apart from a significant improvement in the determination of the intrinsic line
widths and centroids, a major advantage of a higher spectral resolution is also
that contamination due to blending by lines from other ions can be severely
reduced.  Note that our above simulation was done for a slab consisting of pure
\ion{O}{viii}, but in reality the spectrum is composed of a mixture of
absorption lines from many ions with different velocity and ionization
structure.

This is illustrated in Fig.~\ref{fig:figreso}, where I compare the original
model spectrum with the model spectrum folded through 1 and 2~eV instrumental
resolution.  It is clear that for both instrumental resolutions the velocity
fine structure of the absorption lines cannot be resolved.  This is currently a
great advantage of the UV band, where instruments like  FUSE or the STIS
instrument of HST are able to provide very high spectral energy resolution (down
to the 10~km/s scale).  In fact, STIS observations served as the basis for our
velocity model applied in this figure.  Nevertheless, it is evident from this
figure that a 1~eV spectral resolution is a significant improvement over a 2~eV
spectral resolution.  For example, the \ion{O}{viii} Ly$\gamma$ line at
15.18~\AA\ (observed here at 15.4~\AA\ due to the cosmological redshift) and the
\ion{Fe}{xvii} line at 15.27~\AA\ (observed at 15.5~\AA) are not separated with
2~eV resolution but can be distinguished clearly at 1~eV resolution.  With 2~eV
resolution, it would be difficult to see if it is a blend of two narrower
components, or for example a single broad, red/blueshifted line.  The same holds
for separating the \ion{Fe}{xv} and \ion{Fe}{xvi} lines around 15.8~\AA.

Also near the \ion{O}{vii} triplet a gain in spectral resolution is very
important.  With 2~eV resolution the intercombination line of \ion{O}{vii}
cannot be measured, and also the important \ion{O}{vi} inner shell absorption
line, observed at 22.35~\AA\, is only visible with 1~eV resolution, since it is
so close to the strong forbidden emission line.  This \ion{O}{vi} line, as well
as similar inner shell absorption lines from \ion{O}{i} -- \ion{O}{v} at
slightly longer wavelengths, are very important, as they cover a broad range of
ionization parameter with ions from the same chemical element, thus avoiding
modeling difficulties due to unknown elemental abundances.  Although the inner
shell Fe-lines (the UTA mentioned before) around 15--17~\AA\ also cover this
range of ionization parameter, the analysis of these lines is much more
complicated due to the strong blending, while the oxygen lines are in a
relatively clean part of the spectrum (the 21--24~\AA\ range).

\section{Time variability}

\begin{figure}
\centerline{\psfig{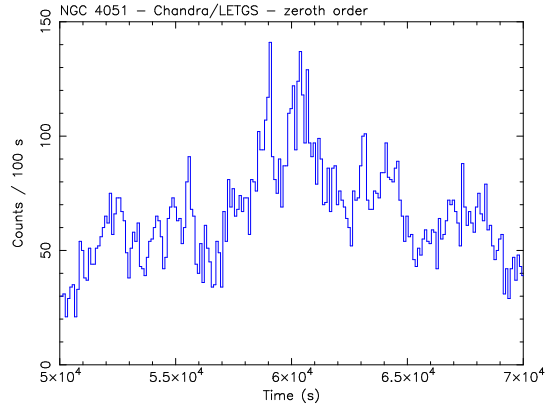}}
\caption[]{Chandra LETGS light curve in first order of NGC~4051. Time
bins: 100~s.}
\label{fig:4051}
\end{figure}

The large effective area of XEUS makes it ideal to study rapid time variability.
For the high luminosity AGN variability typically occurs on time scales of days
or longer, so for those sources variability studies are limited by the finite
exposure time available for each target.  However, for low luminosity AGN the
intrinsic variability time scales can be much faster.  An example is given in
Fig.~\ref{fig:4051}.  This AGN shows rapid variability, down to the time scale
of 100~s.

\begin{figure}
\centerline{\psfig{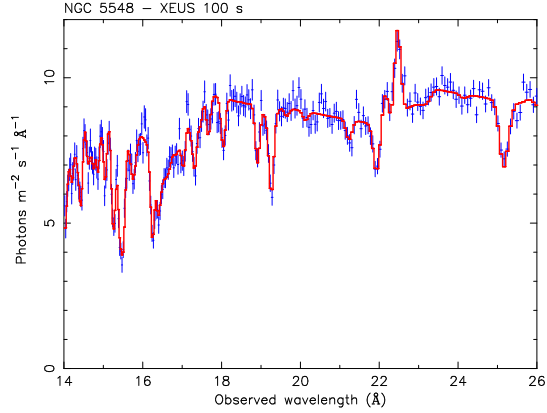}}
\caption[]{Simulated spectrum for NGC~5548 with a 100~s integration time.}
\label{fig:time100}
\end{figure}

With the current generation of grating spectrometers it is not possible to
obtain high quality spectra with such a short integration time on the minute
time scale.  However with the effective area of XEUS this is not a problem.
This is illustrated in Fig.~\ref{fig:time100}.  XEUS is able to obtain high S/N
spectra for bright Seyfert 1 galaxies in just 100~s integration time.  This
allows a study of the intrinsic variations of the accretion disk spectrum on
this time scale.  For example, if an AGN has strong relativistic oxygen and
nitrogen lines, and it varies rapidly on this time scale, all kinds of
reverberation methods can be applied and the geometry of the innermost part of
the accretion disk can be well constrained.

\begin{figure}
\centerline{\psfig{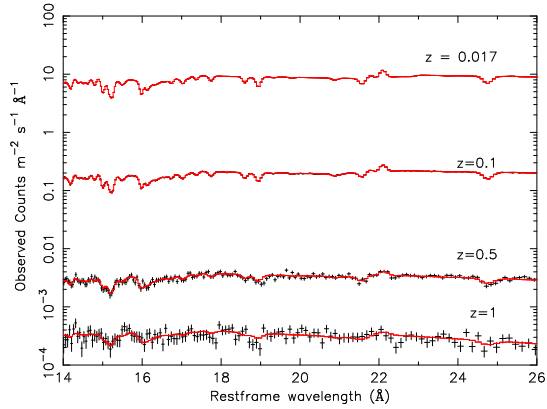}}
\caption[]{Simulated spectra for NGC~5548 with 40~ks exposure time,
for redshifts of 0.017, 0.1, 0.5 and 1 (from top to bottom). The y-axis 
represents the fluxed spectrum in the {\it observers} frame, however for
easy of comparison  the x-axis has been shifted to the rest frame wavelength.
For the lowest redshift, the error bars on the spectrum are barely visible.
}
\label{fig:zplot}
\end{figure}

The large effective area of XEUS is of course not only excellent for high time
resolution observations of bright and nearby AGN, it also will allow to obtain
good spectra of distant AGN.  This is illustrated in Fig.~\ref{fig:zplot}.  In
that figure NGC~5548 is put at redshifts of 0.017 (its true redshift), 0.1, 0.5
and 1.  Good quality spectra within a reasonable net exposure time (half a day)
is possible up to $z=0.5$.  At $z=1$, the spectrum becomes too noisy at the
highest spectral resolution, and also due to the cosmological redshift the
spectral resolution degrades, making it harder to disentangle the narrow
spectral components.  For example, for $z=1$, the \ion{O}{vii} triplet cannot be
resolved, both due to the low source flux and the factor of 2 poorer spectral
resolution, as compared to the case of $z=0.017$.  Of course, for higher
luminosity AGN good spectra can be obtained out to larger redshifts.

\begin{acknowledgements}
SRON is supported
financially by NWO, the Netherlands Organization for Scientific
Research. 
\end{acknowledgements}


\begin{thebibliography}{}
\bibitem[2001]{branduardi}
  Branduardi-Raymont, G., Sako, M., Kahn, S.M., et al., 2001, A\&A 365, L140
\bibitem[2000]{kaastra2000}
  Kaastra, J.S., Mewe, R., Liedahl, D.A., Komossa, S., \& Brinkman, A.C.,
  2000, A\&A 354, L83
\bibitem[2002a]{kaastra2002a}
  Kaastra, J.S., Steenbrugge, K.C., Raassen, A.J.J., et al., 2002a, A\&A
  386, 427
\bibitem[2002b]{kaastra2002spex}
  Kaastra, J.S., Mewe, R., \& Raasen, A.J.J. 2002b, in
  New Visions of the X-ray Universe in the XMM-Newton and Chandra Era,
  ed. F.A. Jansen, ESA, in press
\bibitem[2001]{lee}
  Lee, J.C., Ogle, P.M., Canizares, C.R., et al., 2001, ApJ 554, L13
\bibitem[2001]{sako}
   Sako, M., Kahn, S.M., Behar, E., et al. 2001, A\&A 365, L168
\bibitem[2001]{yaqoob}
  Yaqoob, T., George, I.M., Nandra, K., et al., 2001, ApJ 546, 759
\end{thebibliography}
\end{document}